\documentclass[useAMS, usenatbib, usegraphicx, a4paper]{mn2e}

\usepackage[fleqn]{amsmath} 
\usepackage{amsfonts} 
\usepackage{amstext}
\usepackage{amssymb} 
\usepackage[colorlinks=true, citecolor=blue, linkcolor=blue, breaklinks=true]{hyperref}

\newcommand{\hi } {{\rm H}\,{\small\rm I} \,}

\newcommand{\cvg}{$d_{R}V(0)\;$}

\title[A scaling-relation for disc galaxies]{A scaling-relation for disc galaxies: circular-velocity gradient \textit{versus} central surface brightness}
\author[F. Lelli et al.]{Federico Lelli$^{1}$\thanks{E-mail: lelli@astro.rug.nl},
Filippo Fraternali$^{1, 2}$, Marc Verheijen$^{1}$ \\
$^{1}$Kapteyn Astronomical Institute, University of Groningen, Postbus 800, 9700 AV, Groningen, The Netherlands \\
$^{2}$Department of Physics and Astronomy, University of Bologna, via Berti Pichat 6/2, 40127, Bologna, Italy
}

\begin{document}

\date{}

\maketitle

\begin{abstract}
For disc galaxies, a close relation exists between the distribution of light
and the shape of the rotation curve. We quantify this relation by measuring the
inner circular-velocity gradient $d_{R}V(0)$ for spiral and irregular galaxies
with high-quality rotation curves. We find that $d_{R}V(0)$ correlates with
the central surface brightness $\mu_{0}$ over more than two orders of magnitude
in $d_{R}V(0)$ and four orders of magnitudes in $\mu_{0}$. This is a
scaling-relation for disc galaxies. It shows that the central stellar density
of a galaxy closely relates to the inner shape of the potential well, also
for low-luminosity and low-surface-brightness galaxies that are expected
to be dominated by dark matter.
\end{abstract}


\section{Introduction}\label{sec:intro}

Scaling-relations are an ideal tool to investigate the structure, the formation,
and the evolution of galaxies. For disc galaxies, the Tully-Fisher (TF) relation
\citep{Tully1977} is one of the best-studied scaling laws. It was originally
proposed as a correlation between the absolute magnitude of a galaxy and the
width of its global \hi line profile. It is now clear that the fundamental
relation is between the total baryonic mass of the galaxy and the circular
velocity along the flat part of the outer rotation curve ($V_{\rm flat}$),
thought to be set by the dark matter (DM) halo \citep[e.g.][]{McGaugh2000,
Verheijen2001b, Noordermeer2007b}.

While $V_{\rm flat}$ is related to the total dynamical mass of a galaxy,
the inner shape of the rotation curve provides information on the steepness
of the potential well. For disc galaxies (Sb and later types), the rotation
curve is generally described by an inner rising part (nearly solid-body)
and an outer flat part \citep[e.g.][]{Bosma1981, Begeman1987, Swaters2009}. 
For bulge-dominated galaxies, instead, the rotation curve shows a very fast 
rise in the center, often followed by a decline and the flattening in the
outer parts \citep[e.g.][] {Casertano1991, Noordermeer2007}. 

The relation between the optical properties of a galaxy and the shape of
its rotation curve has been debated for many years \citep[e.g.][]{Rubin1985,
Corradi1990, Persic1991}. Several authors pointed out that the shape of
the luminosity profile and the shape of the rotation curve are closely
related \citep{Kent1987, Casertano1991, Broeils1992, Sancisi2004, Swaters2009}.
In particular, \citet{deBlok1996} and \citet{Tully1997} compared the
properties of two \textit{disc} galaxies on the same position of the TF
relation but with different central surface brightness, and found that
high-surface-brightness (HSB) galaxies have steeply-rising rotation curves
compared to low-surface-brightness (LSB) ones. Thus, for a given total
luminosity (or $V_{\rm flat}$), an exponential light distribution with
a shorter scale-length corresponds to a steeper potential well
(see also \citealt{Amorisco2010}). For HSB spirals, maximum-disc
solutions can explain the dynamics in the central regions with
reasonable values of the stellar mass-to-light ratio $M_{*}/L$
\citep[e.g.][]{vanAlbada1986, Palunas2000}, suggesting that either
baryons dominate the gravitational potential or dark matter closely
follows the distribution of light. \citet{Garrido2005} also found
a clear trend between the inner slope of the rotation curve and the
central surface brightness of 18 HSB spiral galaxies.
For LSB galaxies, maximum-disc solutions can reproduce the inner parts of the
rotation curves, but they often require high values of $M_{*}/L$ that cannot be
explained by stellar population models \citep[e.g.][]{deBlok2001, Swaters2011},
leading to the interpretation that LSB galaxies are dominated by DM at all
radii. Finally, galaxies with a central ``light excess'' with respect to the
exponential disc (e.g. a bulge) show a corresponding ``velocity excess'' in
the inner parts of the rotation curve \citep[e.g.][]{Marquez1999, Swaters2009}.
This is usually referred to as the ``Renzo's rule'' \citep{Sancisi2004}:
for any feature in the luminosity profile of a galaxy there is a corresponding
feature in the rotation curve, and vice versa.

\begin{table*}
\caption{Galaxy sample. The entire table is published in the on-line version of the journal.
The last column provides references for the distance, the surface photometry, and the
rotation curves: a) \citet{Jacobs2009}; b) \citet{Swaters2002b}; c) \citet{Swaters2009};
d) \citet{Saha2006}; e) \citet{Kent1987}; f) \citet{Begeman1987}; g) \citet{Tully2009}; 
h) \citet{Kent1987}; i) \citet{deBlok2008}; j) \citet{Tully1988}; k) \citet{Noordermeer2007a};
l) \citet{Noordermeer2007}. See Sect.~\ref{sec:Data} for details.}
\centering
\begin{tabular}{l c c c c c c c c c c c}
\hline
Name     & Type & Dist & Method & $i$       &$\mu_{0, R}$& $V_{\rm{max}}$ & $d_{R}V(0)$     & $R_{90}$ & $m$& $\chi_{\nu}^{2}$& Ref. \\
         &      & Mpc  &        & $^{\circ}$& mag/$''^{2}$   & km/s           & km/s/kpc    & kpc      &    &           & \\
(1)      & (2)  & (3)  &(4)     & (5)       & (6)            & (7)            & (8)         &(9)       &(10)&(11)       &(12)  \\
\hline
UGC 7559 & IBm  & 5.0$\pm$0.2 & TRGB & 61$\pm$3& 23.6$\pm$0.2 & 33$\pm$3      & 18$\pm$1 & 1.8      & 1   & 0.66     & a, b, c\\
NGC 3198 & SBc  & 14.5$\pm$2.0& Ceph & 71$\pm$3& 20.7$\pm$0.3 & 157$\pm$2     & 52$\pm$8 & 5.3      & 2   & 0.88     & d, e, f\\
NGC 5055 & Sbc  & 7.9$\pm$1.3 & TF   & 59$\pm$3& 18.5$\pm$1.1 & 212$\pm$8     & 418$\pm$72   & 1.7  & 3   & 0.57     & g, h, i\\
UGC 11670& S0/a & 14.2$\pm$5.2& TF   & 70$\pm$3& 15.4$\pm$0.4 & 191$\pm$7     & 3117$\pm$1180& 0.8  & 5   & 0.33     & j, k, l\\
\hline
\end{tabular}
\label{tab:sample}
\end{table*}

In this Letter, we show that the inner circular-velocity gradient
\cvg of a galaxy strongly correlates with the central surface brightness
$\mu_{0}$ over more than two orders of magnitude in \cvg and four
orders of magnitude in $\mu_{0}$, thereby extending and firmly
establishing the correlation hinted at by Fig.~8 of \citet{Garrido2005}.
This is a scaling-relation for disc galaxies. We discuss the
implications of this relation for the stellar and DM properties
of galaxies.

\section{Data Analysis}\label{sec:Data}

\subsection{The circular-velocity gradient}\label{sec:CVG}

We define the circular-velocity gradient \cvg as the inner slope of a galaxy
rotation curve, i.e.\ $dV/dR$ for $R \to 0$. \cvg can be estimated if the
rising part of the rotation curve is well-sampled, but this requires high-quality
data and a careful modelling of the gas kinematics in the inner parts.
To minimize the uncertainties, we use four samples of galaxies with high-quality
rotation curves: \citet{Swaters2009} (hereafter S09), \citet{deBlok2008}
(hereafter dB08 or THINGS), \citet{Verheijen2001a} (hereafter VS01), and
\citet{Begeman1987} (hereafter B87). The rotation curves were derived using
interferometric \hi observations and corrected for beam-smearing effects.
We select only galaxies viewed at inclination angles $i$ between 40$^{\circ}$
and 80$^{\circ}$, as the rotation velocities of face-on discs require a large
correction for $i$, while the observed rotation curves of edge-on discs may be
affected by unseen holes in the central \hi distribution. We projected each
derived rotation curve onto the corresponding position-velocity diagram to
verify that they have been properly determined. We exclude the galaxies
from S09 and VS01 with low-quality data ($q>2$, see S09). The spiral
NGC~3521 (from dB08) is close to edge-on in the inner regions and we
neglect the two innermost velocity points.
Five galaxies are in common between B87 and dB08. We use the new rotation
curves from THINGS except for two galaxies (NGC~2903 and NGC~3198), as the
inner parts of their rotation curves are better traced by B87, who applied
a careful beam-smearing correction (see Figs.~9 and 12 of dB08). The sample
of S09 has one object (NGC~2366) in common with dB08 and another one (UGC~6446)
in common with VS01; we use the rotation curves from S09. In all these cases
the differences in \cvg are, however, within a factor two. The final sample
comprises 63 galaxies with morphological types ranging from Sab to Sd/Im.

To this high-quality sample, we add 11 rotation curves of S0/Sa galaxies 
(with $40^{\circ} \leq i \leq 80^{\circ}$) from \citet{Noordermeer2007}
(hereafter N07). Since early-type galaxies usually lack \hi emission in
their central regions, the rotation curves were derived combining H$\alpha$
long-slit spectroscopy (for the inner parts) with \hi observations (for
the outer parts). We exclude UGC~12043 as the H$\alpha$ observations
of this galaxy have very low velocity resolution. The values of \cvg
for early-type galaxies are more uncertain than those for late-type galaxies.

We derive \cvg by fitting the inner rotation curve with a polynomial function
of the form
\begin{equation}
V(R) = \sum_{\rm n=1}^{m} a_{\rm n} \times R^{n}
\end{equation}
and consider the linear term $a_{\rm 1} = \lim_{R \to 0} dV/dR = d_{\rm{R}}V(0)$.
The fit is error-weighted and constrained to pass through $V=0$ at $R=0$. The value
of $a_{\rm 1}$ depends on i) the radial range used in the fit, and ii) the order
of the polynomial $m$. We define $R_{90}$ as the radius where the rotation curve
reaches 90$\%$ of its maximum velocity, and fit only the points within $R_{90}$.
This choice allows us to maximize the number of points along the rising part
of the rotation curve without including points along the flat part.
Rotation curves with less then 3 points within $R_{90}$ are excluded,
as they are not well-resolved in the inner parts. 16 galaxies from the
high-quality sample and 3 galaxies from N07 are excluded by this criterion,
thus the high-quality and total samples reduce to 47 and 55 objects,
respectively. For a pure exponential disc with scale-length~$R_{\rm{d}}$,
$R_{90} \simeq 1.2 R_{\rm{d}}$. Thus, the first fitted point of the
rotation curve is typically at $R\lesssim 0.4 R_{\rm{d}}$ and the
derived value of $d_{\rm{R}}V(0)$ is representative of the
innermost galaxy regions that are accessible by the available rotation
curves. To derive the best-fit order of the polynomial, we proceed as
follows. We start with a linear fit ($m=1$) and progressively increase $m$
until the reduced-$\chi^{2}$ ($\chi_{\nu}^{2}$) approaches 1.
In practice, we minimize the function $P_{\chi}(\chi^2; \nu)-0.5$,
where $P_{\chi}(\chi^2; \nu)$ is the integral probability of $\chi^2$
and $\nu$ is the number of degrees of freedom; the procedure is
halted in case $\chi_{\nu}^2$ would drop below 1.
Visual inspection showed that this method works better than the
F-test \citep[e.g.][]{BR03}, that in some cases returns high values
of $m$ and thus increases the number of free parameters in the fit.

We test our automatic procedure on a set of model rotation curves,
calculated by summing the contributions of a disc, a bulge, and a
DM halo. We add typical errors to the velocity points ($\sim$5
km~s$^{-1}$) and try several spatial samplings. We find that,
even if the rotation curve is poorly-sampled ($\sim$5 points within
$R_{90}$), the actual value of $d_{\rm{R}}V(0)$ can be recovered with
a error of $\sim$30$\%$. However, if the rotation curve has an inner
``bump'' (due to a compact bulge), $d_{\rm{R}}V(0)$ may be
under-estimated by a factor $\sim$2.

\begin{figure}
\centering
\includegraphics[width=8.2 cm, angle = -90]{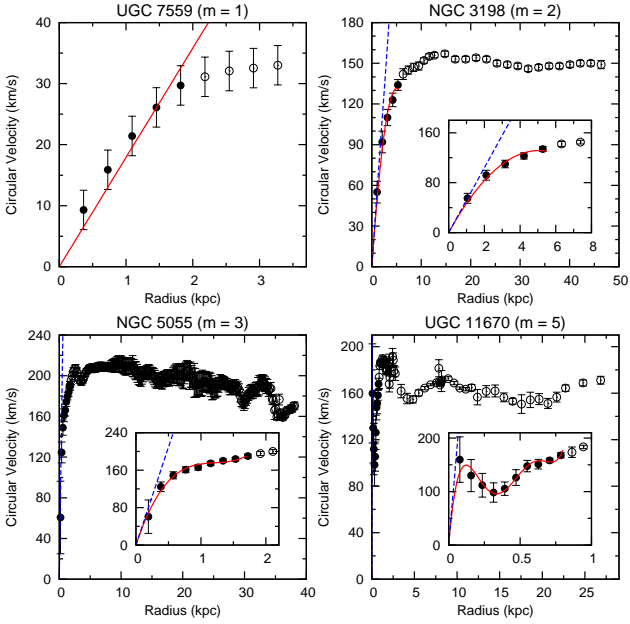}
\caption{Results of the polynomial-fit for four representative galaxies.
Filled circles show the points of the rotation curve within $R_{90}$,
while open circles show the points excluded in the fit. The solid, red
line shows the fitted polynomial function, while the dashed, blue line
shows its linear term. The order $m$ of the polynomial is indicated.
See Sect.~\ref{sec:CVG} for details.
}
\label{fig:CVG}
\end{figure}

Figure \ref{fig:CVG} shows the results for four representative galaxies that
require polynomial fits of different orders. Late-type galaxies (Sb to Im)
are generally well-fitted by polynomials with $m=1$ (e.g. UGC\,7559) or $m=2$
(e.g. NGC\,3198), but several cases do require $m\geq3$ (e.g. NGC\,5055).
Early-type galaxies (S0/Sa) often require high-order polynomials ($m \geq 4$,
e.g. UGC\,11670), as their rotation curves may have complex shapes characterized
by a steeply-rising part followed by a decline and a second rise. For some
bulge-dominated galaxies from N07, the value of \cvg is rather uncertain,
since there may be no data points in the inner radial range where the linear
term $a_{1}$ is representative of the rotation curve (see Fig.~\ref{fig:CVG},
bottom-right).
Table~\ref{tab:sample} provides the fit results for the galaxies in
Fig.~\ref{fig:CVG}. The results for the entire galaxy sample are
provided in the on-line version of this Table.

The error $\delta_{d_{\rm{R}}V(0)}$ on \cvg is estimated as
\begin{equation}\label{eq:errGrad}
\delta_{d_{\rm{R}}V(0)} =\sqrt{  \delta_{a_{1}}^{2} +
                      \bigg( d_{\rm{R}}V(0) \dfrac{\delta_{i}}{\tan(i)}\bigg)^{2} +
		       \bigg( d_{\rm{R}}V(0) \dfrac{\delta_{\rm{D}}}{D}\bigg)^{2} }
\end{equation}
where $\delta_{a_{1}}$ is the nominal error on the fitted linear term $a_{1}$,
$\delta_{i}$ is the error on the disc inclination $i$, and $\delta_{\rm{D}}$ is
the error on the galaxy distance $D$. $\delta_{\rm{D}}$ typically gives a negligible
contribution for galaxies with distances derived using the tip of the red giant
branch (TRGB) and/or Cepheids (Ceph), whereas it can dominate the error budget for
galaxies with distances estimated from the TF relation or the Hubble flow.

\subsection{The central surface brightness}

For the high-quality sample of disc-dominated galaxies, we consider two
ways to estimate the central surface brightness: i) the disc central
surface brightness $\mu_{\rm{d}}$, obtained from an exponential fit to
the outer parts of the luminosity profile; and ii) the observed central
surface brightness $\mu_{0}$, obtained from a linear extrapolation of
the luminosity profile in the inner few arcseconds to $R=0$ (see \citealt{Swaters2002b}).
$\mu_{0}$ takes into account possible deviations from a pure exponential disc.
This may carry valuable information on the mass distribution, e.g. if a
pseudo-bulge/bar is present, but may also reflect variations in the stellar
populations and/or in the internal extinction, e.g. if the star-formation
activity is enhanced in the central parts. We use the observed central
surface brightness $\mu_{0}$. Since we are considering disc-dominated
galaxies (Sb and later types), we correct $\mu_{0}$ for inclination;
we assume an optically-thin disc. Given the ambiguity in using either
$\mu_{0}$ or $\mu_{\rm{d}}$, we include the difference
$\Delta \mu = \mu_{\rm{d}} - \mu_{0}$ in the error $\delta_{\mu_{0}}$.
This is estimated as
\begin{equation}\label{eq:SBerr}
\delta_{\mu_{0}} = \sqrt{(\Delta \mu/2)^{2} + [2.5 \log(e)\tan(i)\delta_{i}]^{2}}.
\end{equation}

For the galaxies from S09, we use the values listed in Table A.5 of \citet{Swaters2002b}
(Harris $R$-band). For the galaxies from VS01, we use the surface photometry from
\citet{Tully1996} (Cousins $R$-band). For the galaxies from dB08 and B87, we use
the surface photometry from three different sources (in order of preference):
\citet{Swaters2002b} (Harris $R$-band), \citet{Kent1987} ($r$-band), and \citet{Munoz2009}
(Harris $R$-band or $r'$-band). The optical filters are comparable, but there can
be systematic differences of $\sim$0.1 mag (within the typical errors). Two galaxies
from dB08 (NGC~925 and NGC~7793) and one galaxy from B87 (NGC~5371) have no $R$-band
photometry available and have been excluded, reducing the total sample to 52 objects.

The S0/Sa galaxies from N07 require a different approach, because i) the surface
brightness rapidly increases in the central regions due to the presence of a
dominant bulge; and ii) several galaxies are at large distances ($\gtrsim$30 Mpc),
thus the linear resolution of the optical observations is not very high
($\gtrsim$150 kpc). For these galaxies, \citet{Noordermeer2007a} provide
the $R$-band disc central surface brightness $\mu_{\rm{d}}$, extrapolated
from an exponential fit and corrected for $i$, and the bulge central surface
brightness $\mu_{\rm{b}}$, extrapolated from a Sersic fit to the inner parts
after subtracting the disc contribution. We estimate $\mu_{0}$ by summing the
contributions of $\mu_{\rm{d}}$ and $\mu_{\rm{b}}$; the latter value is
\textit{not} corrected for $i$ as the bulge is assumed to be spherical.
The errors are given by Eq.~\ref{eq:SBerr}, where $\Delta \mu$ is now the
difference between $\mu_{0}$ and the innermost value of $\mu$ observed.

\begin{figure*}
\centering
\includegraphics[width=7.8cm, angle=-90]{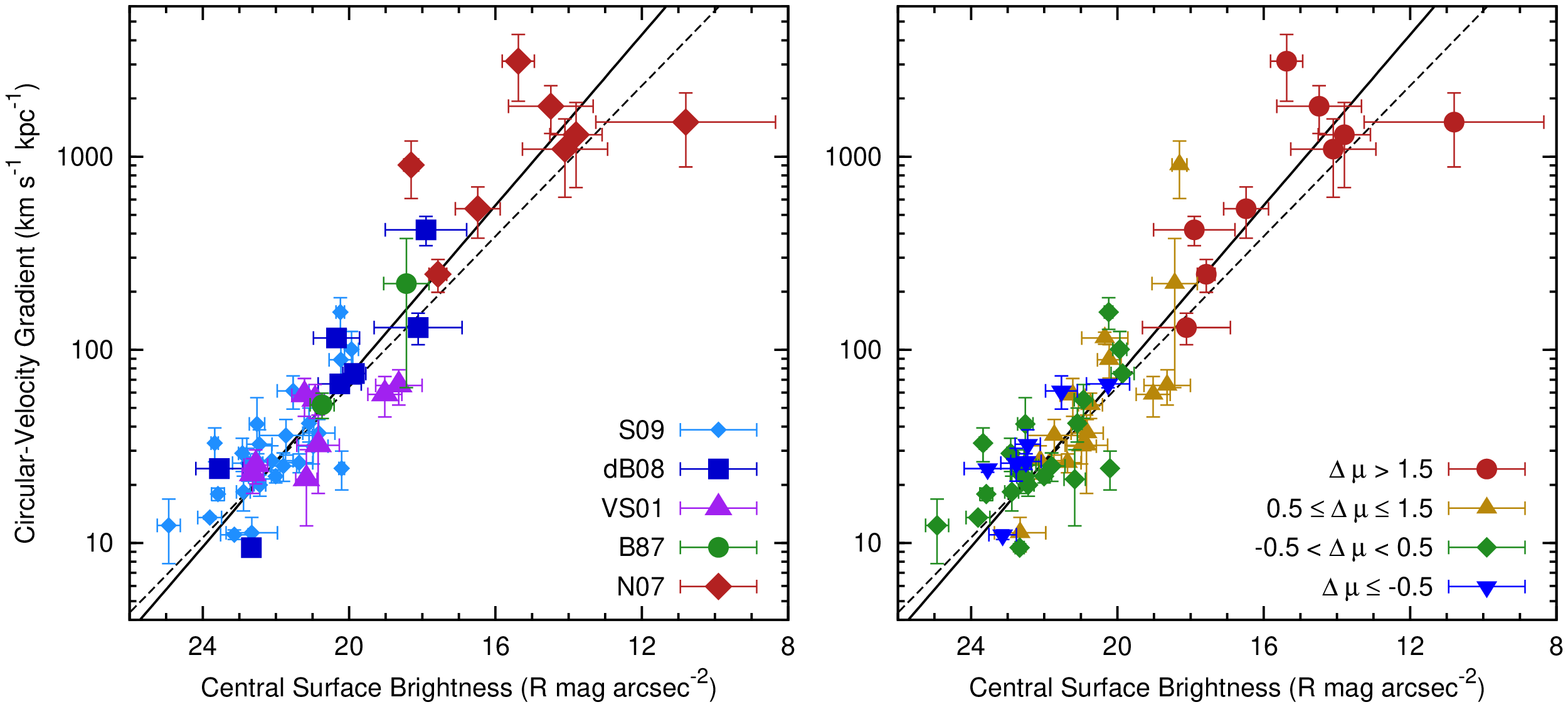}
\caption{The circular-velocity gradient versus the central surface brightness.
The solid and dashed lines show a liner fit to the data-points for the total and
high-quality samples, respectively. \textit{Left:} galaxies coded by the reference
for the rotation curve
(S09: \citealt{Swaters2009}, dB08: \citealt{deBlok2008}, VS01: \citealt{Verheijen2001a},
B87: \citealt{Begeman1987}, N07: \citealt{Noordermeer2007}).
\textit{Right:} galaxies coded by the value of $\Delta \mu = \mu_{\rm{d}} - \mu_{0}$
(in R mag arcsec$^{-2}$), that quantifies the ``light excess'' over an exponential profile.}

\label{fig:plot}
\end{figure*}
\section{The \cvg$-\,\mu_0$ scaling relation}\label{sec:plot}

In Fig.~\ref{fig:plot} (left), we plot $\mu_{0}$ against \cvg for
the total sample of 52 galaxies. There is a clear, striking relation.
A linear, error-weighted fit to the data yields
\begin{equation}
 \log [d_{R}V(0)] = (-0.22\pm 0.02) \, \mu_{0} + (6.28\pm 0.40).
\end{equation}
As discussed in Sect.~\ref{sec:Data}, the values of \cvg for the S0/Sa
galaxies from N07 are uncertain. However, it is clear that these
bulge-dominated galaxies follow the same trend defined by disc-dominated ones.
Fig.~\ref{fig:plot} (left) also shows a linear fit excluding the objects
from N07 (dashed line). This gives only slightly different values of the
slope ($-0.19\pm0.03$) and the intersect ($5.70\pm0.57$). 
Considering the different types of galaxies and the uncertainties involved,
the relation shown in Fig.~\ref{fig:plot} is remarkably tight and extended,
spanning more than two orders of magnitude in \cvg and four orders of
magnitude in $\mu_{0}$.

The values of the slope and the intersect are likely more uncertain than
the formal errors, due to several effects in the determination of \cvg
and $\mu_{0}$. Possible concerns are i) the different linear resolutions
(in kpc) of the \hi and optical observations, and ii) the effects on
$\mu_{0}$ of internal extinction, recent star-formation, and/or a LINER core.
We performed several fits using different methods to estimate $\mu_{0}$
and $d_R V(0)$, such as calculating $V/R$ at the innermost point of the
rotation curve. We obtained slopes always between $-0.25$ and  $-0.15$,
and we think that the actual slope must be constrained between these values. 

The scatter around the relation is largely due to observational uncertainties
on $d_{R}V(0)$. Major sources of uncertainties are i) the galaxy distance,
ii) the inclination, and iii) the innermost points of the rotation curve (see
Eq.~\ref{eq:errGrad}). However, part of the scatter is likely to be intrinsic
and due to differences in the 3-dimensional (3D) distribution of baryons
and in the structural component that defines $\mu_{0}$ (a disc, a bulge,
a bar, or a nuclear star cluster).

To investigate the role played by different structural components, in
Fig.~\ref{fig:plot} (right) we plot the same data-points coding the galaxies
by the value of $\Delta \mu = \mu_{\rm{d}} - \mu_{0}$. This quantifies the
deviation from an exponential law in the inner parts of the luminosity profile
(in R mag arcsec$^{-2}$). We distinguish between four cases: i) galaxies
dominated by a bulge ($\Delta \mu > 1.5$), ii) galaxies with a small
central concentration of light ($0.5\leq \Delta \mu \leq 1.5$) like
a pseudo-bulge or a bar, iii) galaxies with an exponential disc
($-0.5< \Delta \mu < 0.5$), and iv) galaxies with a central light depression
($\Delta \mu \leq -0.5$). The upper-right end of the relation ($\mu_{0} \gtrsim 18$
mag~arcsec$^{-2}$) is populated by bulge-dominated galaxies. It is clear that,
for these galaxies, the use of $\mu_{\rm{d}}$ instead of $\mu_{0}$ would
shift them away from the relation, as $\mu_{\rm{d}} \lesssim$ 19-20
R~mag~arcsec$^{-2}$ (the ``Freeman value'', \citealt{Freeman1970}). On the
lower-left end of the relation, instead, one can find both pure exponential
discs and galaxies with central light concentrations/depressions. For these
disc-dominated galaxies, the use of $\mu_{\rm{d}}$ instead of $\mu_{0}$ would
still lead to a correlation, but this would have a steeper slope ($\sim$-0.25).
For galaxies with similar values of $\mu_{0}$, \cvg does not seem to depend
on the detailed shape of the luminosity profile (simple exponential
or with a central light depression/concentration).

\section{Discussion}\label{sec:disc}

The correlation between the central surface brightness $\mu_{0}$ and the
circular-velocity gradient \cvg implies that there is a close link between
the stellar density and the gravitational potential in the central parts
of galaxies. This holds for both HSB and LSB objects, covering a wide
range of masses and asymptotic velocities ($20\lesssim V_{\rm flat}
\lesssim 300$ km~s$^{-1}$). 

The relation between the distribution of light and the distribution of mass
has been extensively discussed in the past (see \citealt{Sancisi2004} and
references therein). However, only few attempts have been made to parametrize
this relation, notably by \citet{Swaters2009} (S09). Fig.~10 of S09 plots
the logaritmic-slope between 1 and 2 disc
scale-lenghts $S_{1,2}=\log[V(2h)/V(h)]/\log(2)$ versus the ``light
excess'' with respect to an exponential disc $\Delta \mu = \mu_{\rm d} -
\mu_{0}$. It shows that a ``light excess'' (a bulge-like component)
corresponds to a ``velocity excess'' in the rotation curve with respect
to the expectations for the underlying exponential disc. The limitation
of that parametrization is that it does not capture the dynamical difference
between HSB and LSB discs, that are known to have steeply-rising and
slowly-rising rotation curves, respectively \citep[e.g.][]{Tully1997}.
In Fig.~10 of S09, indeed, both HSB and LSB exponential discs have
$\Delta \mu \simeq 0$ and $S_{1,2}\simeq 0.5$. The latter result is
due to the fact that $S_{1,2}$ is, by definition, a scale-invariant
quantity, that is not expected to depend on $\mu_{0}$ or $V_{\rm{max}}$.
In contrast, \cvg measures the inner slope of the rotation curve in
physical units (km~s$^{-1}$~kpc$^{-1}$) and is directly related to
the central dynamical surface density (in $M_{\odot}$~pc$^{-2}$),
providing insights in the underlying physics, as we now discuss.

For a 3D distribution of mass, the rotation velocity $V$ of a test
particle at radius $R$ is given, to a first approximation, by
\begin{equation}
\dfrac{V^{2}}{R} = \alpha \dfrac{G M_{\rm{dyn}}} {R^{2}}
\end{equation}
where $G$ is Newton's constant, $M_{\rm{dyn}} = 4/3 \pi R^{3}
\overline{\rho}_{\rm{dyn}}$ is the dynamical mass within $R$, and
$\alpha$ is a factor that depends on the detailed mass distribution
(for a spherical distribution of mass $\alpha$=1, while for a thin
exponential disk $\alpha \simeq 0.76$ at $R = 0.5 R_{0}$).
For $R \to 0$, we have
\begin{equation}
\dfrac{dV}{dR} = \dfrac{V}{R} = \sqrt{\beta G \rho_{\rm{dyn, 0}}} = \sqrt{\beta G \dfrac{\rho_{\rm{bar, 0}}}{f_{\rm{bar, 0}}}}
\end{equation}
where $\beta = 4/3 \pi \alpha$, $\rho_{\rm{dyn, 0}}$ and $\rho_{\rm{bar, 0}}$ are,
respectively, the central dynamical and baryonic mass densities, and $f_{\rm{bar, 0}}
= \rho_{\rm{bar, 0}} /\rho_{\rm{dyn, 0}}$ is the baryon fraction in the central regions.
Note that $f_{\rm{bar, 0}}$ may strongly differ from the ``cosmic'' baryon fraction,
and can vary widely from galaxy to galaxy, depending on the formation and
evolution history. Observationally, we measure $\mu_{0}$ which is related
to $\rho_{\rm{bar, 0}}$ by
\begin{equation}
 \mu_{0} = -2.5 \log [\rho_{\rm{bar, 0}} \, \Delta z \, (M_{\rm{bar}}/L)^{-1} ]
\end{equation}
where $\Delta z$ is the typical thickness of the stellar component (either a disk
or a bulge) and $M_{\rm{bar}}/L$ is the baryonic mass-to-light ratio, including
molecules and other {\it dark} baryonic components. Thus, we expect the following
relation
\begin{equation}\label{eq:Newt}
 \log [d_R V(0)] = -0.2 \, \mu_{0} + 0.5 \log \bigg(\beta G \dfrac{M_{\rm{bar}}/L}{\Delta z \, f_{\rm{bar, 0}}} \bigg).
\end{equation}

In Sect.~\ref{sec:plot}, we mentioned that the slope of our relation is not
well-determined due to several uncertainties in the measurements of \cvg and
$\mu_{0}$. However, it is consistent with $-0.2$ and can be constrained between
$-0.15$ and $-0.25$. Were the slope exactly $-0.2$, the second term of
Eq.~\ref{eq:Newt} would be a constant, implying a puzzling fine-tuning between
the 3D distribution of baryons ($\beta$ and $\Delta z$), the baryonic
mass-to-light ratio ($M_{\rm{bar}}/L$), and the dark matter content ($f_{\rm{bar, 0}}$).

Despite the uncertain value of the slope, the results presented here show a
clear relation between the central stellar density in a galaxy and the
steepness of the potential well (see also \citealt{Sancisi2004, Swaters2011}).
This implies a close link between the density of the baryons, regulated by
gas accretion, star-formation, and feedback mechanisms, and the central
density of the DM halo, together shaping the inner potential well. This
may represent a challenge for models of galaxy formation and evolution.
Future observational studies may help to better constrain the slope of
the relation, while theoretical work should aim to understand its origin.

\section{Conclusions}

We measured the circular-velocity gradient \cvg for a sample of spiral and
irregular galaxies with high-quality rotation curves. We found a linear
relation between $\log[d_{\rm R}V(0)]$ and the central surface brightness
$\mu_{0}$ with a slope of about $-0.2$. This is a scaling-relation for
disc galaxies that holds for objects of very different morphologies, 
luminosities, and sizes, ranging from dwarf irregulars to bulge-dominated
spirals. This relation quantifies the coupling between visible and
dynamical mass in the central parts of galaxies, and shows that the
central stellar density closely relates to the inner shape of the
potential well.\\
\newline
\textit{Acknowledgements:} We are grateful to Renzo Sancisi for stimulating
discussions and insights. We also thank Erwin de Blok and Rob Swaters for
providing us with the high-quality rotation curves.

\renewcommand\bibname{{References}}
\bibliographystyle{mn2e.bst}
\bibliography{LFV.letter.bib}

\end{document}